\documentclass[prl,showpacs,twocolumn,superscriptaddress,floatfix,amsmath]{revtex4}
\usepackage[latin1]{inputenc}
\usepackage[english]{babel}
\usepackage{graphicx}
\usepackage{psfrag}
\usepackage{amssymb}
\usepackage{amsmath}
\usepackage{amscd}
\usepackage{eucal}
\usepackage{color}
\usepackage{bm}
\newcommand{\bi}{\bibitem}
\newcommand{\ignore}[1]{\relax}

\usepackage{epstopdf}
\DeclareMathAlphabet{\mathpzc}{OT1}{pzc}{m}{it}

\newcommand{\tEc}{\tau_{\rm E}^{\rm cl}}
\newcommand{\tEo}{\tau_{\rm E}^{\rm op}}

\begin{document}

\title{Dephasing in the semiclassical limit is system--dependent}

\author{ Cyril Petitjean}
\affiliation{D\'epartement de Physique Th\'eorique,
Universit\'e de Gen\`eve, CH-1211 Gen\`eve 4, Switzerland}

\author{Philippe Jacquod}
\affiliation{Physics Department, University of Arizona, 
1118 E. 4$^{\rm th}$ Street, Tucson, AZ 85721, USA}

\author{Robert S.~Whitney}
\affiliation{Institut Laue-Langevin,
6 rue Jules Horowitz, BP 156, 38042 Grenoble, France}

\date{December 5, 2006 -- updated November 21, 2007}
\begin{abstract}
We investigate dephasing in open quantum chaotic
systems in the limit of 
large system size to Fermi wavelength ratio, $L/\lambda_{\rm F} \gg 1$.
We semiclassically calculate the weak 
localization correction $g^{\rm wl}$ to the conductance
for a quantum dot coupled to (i) an external closed dot and (ii)
a dephasing voltage probe.
In addition to the universal algebraic
suppression $g^{\rm wl} \propto (1+\tau_{\rm D}/\tau_\phi)^{-1}$
with the dwell time $\tau_{\rm D}$ through the cavity 
and the dephasing rate $\tau_\phi^{-1}$,
we find an
exponential suppression of weak localization by a factor 
$\propto \exp[-\tilde{\tau}/\tau_\phi]$, with a system-dependent 
$\tilde{\tau}$.
In the dephasing probe model, $\tilde{\tau}$ coincides with 
the Ehrenfest time, $\tilde{\tau} \propto \ln [L/\lambda_{\rm F}]$,
for both perfectly and partially transparent dot-lead couplings. 
In contrast, 
when dephasing occurs due to the coupling to an external dot,
$\tilde{\tau} \propto \ln [L/\xi]$ depends on 
the correlation length $\xi$ of the coupling potential instead of
$\lambda_{\rm F}$. 
\end{abstract}
\maketitle 

{\bf Introduction.} Electronic transport 
in mesoscopic systems exhibits a range of quantum coherent
effects such as weak localization, universal conductance fluctuations
and Aharonov-Bohm effects~\cite{Sto92,Imr97}.
Being intermediate in size between micro- and
macroscopic systems, these systems are
ideal playgrounds to investigate
the quantum-to-classical transition from a microscopic coherent world, where
quantum interference effects prevail, to a macroscopic classical 
world~\cite{Joo03}. Indeed, the disappearance of quantum coherence
in mesoscopic systems as dephasing processes 
set in has been the subject of intensive theoretical~\cite{Alt82,But86,Bar99,
dephasing} and experimental
~\cite{Hui98,Han01,Pie03} studies.
When the temperature is sufficiently low, it is accepted that 
the dominant processes of dephasing are electronic interactions. In disordered
systems, dephasing due
to electron-electron interactions is known to be well modeled
by a classical noise potential~\cite{Alt82}, which 
gives an algebraic suppression of the weak localization correction
to conductance through a diffusive quantum dot, 
\begin{equation}
g^{\rm wl} = g_0^{\rm wl} \big/ (1+\tau_{\rm D}/\tau_\phi).
\label{eq:dephasing-without-Ehrenfest}
\end{equation}
Here, $ g_0^{\rm wl}$ is the weak localization correction 
(in units of $2e^2/h$) without dephasing,
the dephasing time $\tau_\phi$ is given by the noise power,
and  $\tau_{\rm D}$ is the electronic dwell time in the dot.
Eq.~\eqref{eq:dephasing-without-Ehrenfest}
is insensitive to most noise-spectrum details, and holds for 
other noise sources such as
electron-phonon interactions or external microwave fields.

\begin{figure}
\begin{center}
\includegraphics[width=8.cm]{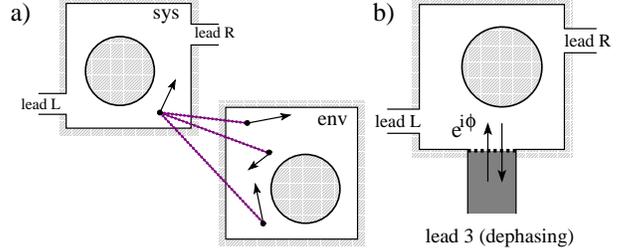} 
\caption{\label{fig:two_models} 
a) (left panel) Schematic of the system-environment model.
The system is an open quantum dot that is coupled to an environment in the 
shape of a second, closed quantum dot. b) (right panel) Schematic of the dephasing lead model.}
\end{center}
\end{figure}

Other, mostly phenomenological models of dephasing have been proposed
to study dephasing in ballistic systems~\cite{But86,dephasing,Bar99}, 
the most popular
of which, perhaps, being the dephasing lead model~\cite{But86,Bar99}. 
A cavity is connected to two external, L (left)
and R (right) leads
of widths $W_{\rm L},W_{\rm R}$.
A third lead of width $W_3$ is connected to the system via a tunnel-barrier of
transparency $\rho$. A voltage is applied to the third lead to ensure that
no current flows through it on average.
A random matrix theory (RMT) treatment of the dephasing lead model
leads to Eq.~(\ref{eq:dephasing-without-Ehrenfest})
with $\tau_{\rm D} = \tau_0L/(W_{\rm L}+W_{\rm R})$
and $\tau_\phi = \tau_0L/(\rho W_3)$, in term of the dot's time
of flight $\tau_0$~\cite{Bar99,Ben97}. 
Thus it is commonly assumed that dephasing is system-independent.
The dephasing lead model is often used phenomenologically 
in contexts where the source of dephasing is unknown.

Our purpose in this article is to revisit dephasing in open chaotic
ballistic systems with a focus on whether dephasing remains system-independent 
in the semiclassical limit of large ratio $L/\lambda_{\rm F}$ 
of the system size to Fermi wavelength. This regime
sees the emergence of a finite Ehrenfest time scale,
$\tEc = \lambda^{-1} \ln [L/\lambda_{\rm F}]$  ($\lambda$ is the 
Lyapunov exponent),
in which case dephasing can lead to an exponential
suppression of weak localization, $\propto \exp[-\tEc/\tau_\phi] \big/
(1+\tau_{\rm D}/\tau_\phi)$~\cite{Ale96}.
Subsequent numerical investigations on the dephasing lead model
support this prediction~\cite{Two04}. 
Here we analytically investigate two different
models of dephasing, and show that the suppression of weak localization 
is strongly system-dependent.
First, we construct a new formalism that incorporates 
the coupling to external degrees of freedom into the scattering approach
to transport. This approach is illustrated by a semiclassical 
calculation of weak localization in the case of an environment modeled
by a capacitively coupled, closed quantum dot. We restrict ourselves 
to the regime of pure dephasing,
where the environment does not alter the classical dynamics of the system.
Second, we provide the first
semiclassical treatment of transport in the dephasing lead model. 
We show that in both cases, the weak localization correction
to conductance is 
\begin{equation}\label{eq:gwl-bothcases}
g^{\rm wl} = g^{\rm wl}_0 \; 
\exp[-\tilde{\tau}/\tau_\phi] \; \big/ (1+\tau_{\rm D}/\tau_\phi),
\end{equation}
where $g^{\rm wl}_0$ is 
the finite-$\tEc$ correction in absence of dephasing.
The time scale $\tilde{\tau}$ is system-dependent.
For the dephasing
lead model, $\tilde{\tau}=\tEc + (1-\rho)\tEo$ in terms of
the transparency $\rho$ of the contacts to the leads, and the open
system Ehrenfest time $\tEo=\lambda^{-1} \ln [W^2/\lambda_{\rm F} L]$.  
This analytic result fits the numerics of Ref.~\cite{Two04}, 
and (up to logarithmic corrections) is in agreement with Ref.~\cite{Ale96}. 
Yet for the system-environment model, 
$\tilde{\tau} = \lambda^{-1}\ln [(L/\xi)^2]$ 
depends on the correlation length $\xi$ of the inter-dot coupling 
potential. We thus conclude that dephasing in the semiclassical limit 
is system-dependent. 

{\bf Transport theory for a system-environment model.}
 In the standard theory of decoherence, one starts with the 
total density matrix $\eta_{\rm tot}$
including both system and environment degrees of freedom~\cite{Joo03}. 
The time-evolution of $\eta_{\rm tot}$ is unitary. 
The observed properties of the system alone are given by 
the reduced density matrix $\eta_{\rm sys} $, obtained
by tracing $\eta_{\rm tot}$ over the environment degrees of freedom. 
This is probability conserving,
${\rm Tr} \; \eta_{\rm sys} =1 $, but renders the time-evolution of 
$\eta_{\rm sys}$ non-unitary. The decoherence time is inferred from 
the decay rate of its off-diagonal matrix elements~\cite{Joo03}. 
We generalize this approach to the scattering theory of transport.

To this end, we consider two coupled chaotic cavities as sketched in Fig.~1a. 
Few-electron double-dot systems similar to 
the one considered here have recently 
been the focus of intense experimental efforts~\cite{Wiel03}.
One of them (the system) is  an open 
quantum dot connected to two external leads. The other one
(the environment) 
is a closed quantum dot, which we model using RMT. 
The two dots are chaotic and 
capacitively coupled. In particular, they do not
exchange particles. We require that $\lambda_{\rm F} \ll W_{\rm L,R} \ll L$, 
so that the number of transport channels satisfies 
$1 \ll N_{\rm L,\,R}  \ll L/\lambda_F$ and the chaotic dynamics inside the 
dot has enough time to develop, $\lambda \tau_{\rm D} \gg 1$. 
Electrons in the leads do not interact with the second dot. 
Inside each cavity the dynamics is generated by chaotic
Hamiltonians $ H_{\rm sys}$ and $H_{\rm env}$. 
We only specify that the capacitative coupling potential, ${\cal U}$,
is smooth, and has magnitude $U$ and correlation length $\xi$.

The environment coupling can be straightforwardly included
in the scattering approach, by writing
the scattering matrix, ${\mathbb S}$, as an integral over time-evolution 
operators.
We then use a bipartite semiclassical propagator
to write the matrix elements of 
${\mathbb S}$ for given initial and final environment positions, 
$({\bf q}_{0},{\bf q} )$, as
\begin{eqnarray} \label{sc-smatrix}
{\mathbb S}_{\rm m n} (  {\bf q}_{0} , {\bf q} ) &=&  
  (2\pi)^{-1}
 \int_{0}^{\infty}\!{\rm d}t \! \int_{\rm L}\! {\rm d} {\bf y}_{\rm 0 } \! \int_{\rm R} \!{\rm d} {\bf y} \,
\left \langle m  \vert {\bf y}  \right\rangle \left\langle {\bf y}_{\rm 0} \vert n \right\rangle
  \,
 \nonumber  \\ &\times& 
 \sum_{\gamma ,\Gamma}
( C_{\gamma} \; C_{ \Gamma})^{\frac{1}{2}} 
 \exp [
 {\it i } 
 \left\{ 
 S_{\gamma} +
  S_{\Gamma }
   +
   {\cal S} _{\gamma ,\Gamma}   \right\}
    ]. \quad
\end{eqnarray}
This is a double sum over classical paths, 
labeled $\gamma$ for  the system and  $\Gamma $ for the environment.  
For pure dephasing, the classical  path $\gamma$ ($\Gamma$) connecting 
${\bf y_{0}}$ (${\bf q_{0}}$) to ${\bf y}$ ($ {\bf q }$) in  the time $t$ is solely determined  by  
$H_{\rm sys} $ ($ H_{\rm env}$). 
The prefactor  $C_{\gamma}C_{\Gamma}$ is the inverse determinant of the 
stability matrix, 
and the exponent contains the non-interacting action integrals, 
$S_{\gamma},S_{\Gamma}$, accumulated along $\gamma$ and $\Gamma $, 
and the interaction term, 
${\cal S}_{\gamma ,\Gamma} =\int_0^{t} \!{\rm d}\tau  {\cal U}[{\bf y}_{\gamma}(\tau), {\bf q}_{\Gamma}(\tau)]$.

Since we assume that particles in the leads do not interact 
with the second cavity, we can write the initial total density matrix as 
$\eta^{(n)} =\eta_{\rm sys}^{(n)} \otimes  \eta_{\rm env}$, with $\eta_{\rm sys}^{(n)} =\vert n \rangle  \langle n\vert$, $n=1,2,... N_{\rm L}$. We
take $ \eta_{\rm env} $ as a random matrix, though
our approach is not restricted to that particular choice.
We define the conductance matrix 
as the following trace over the environment degrees of freedom,
\begin{equation}\label{reducedcond}
g_{mn}^{(r)} =
\big\langle m \big\vert  {\rm Tr}_{\rm env}\big[
{\mathbb S} \;
\eta^{(n)} \;
{\mathbb S}^{\dagger}
 \big] \big\vert m \big \rangle.
\end{equation}
The conductance is then given by $g = \sum_{m,n} g_{mn}^{(r)}$. 
This construction is current conserving,
however the environment-coupling generates decoherence and the 
suppression
of coherent contributions to transport. To see this we now calculate the 
conductance to leading order in the weak localization correction.
 
\begin{figure}
\includegraphics[width=8.0cm]{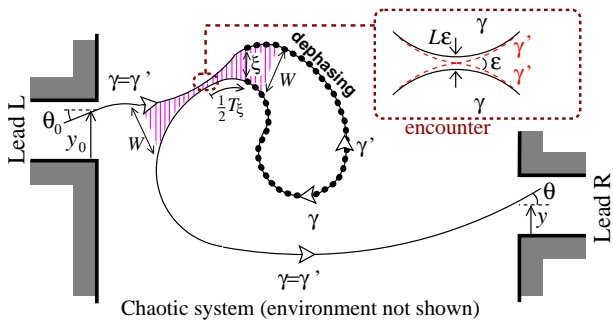} 
\caption{\label{fig:weakloc} 
(color online) A semiclassical contribution to weak localization for 
the system-environment model.
The paths are paired everywhere except at the 
encounter. There one crosses itself at angle $\epsilon$,
while the other does not (going the opposite way around the loop).
Here we show $\xi>\epsilon L$, so the dephasing (dotted path segment)
starts in  the loop ($T_{\xi} >0$). 
}
\end{figure}

We insert Eq.~\eqref{sc-smatrix} into Eq.~\eqref{reducedcond}, 
perform the sum over channel indices with the
semiclassical approximation
$\sum_{n}^{N_{\rm L} } \langle {\bf y}_{0} \vert n \rangle  
\langle n \vert  {\bf y}^{\prime}_{0}  \rangle
\approx \delta( {\bf y}^{\prime}_{0} -{\bf y}_{0})$~\cite{Jac06},
and use the RMT result
$\overline{\langle {\bf q}_{0} \vert \eta_{\rm env} \vert {\bf q}_{0}^{\prime} \rangle }\approx \Omega_{\rm env}^{-1}  \delta({\bf q}_{0}^{\prime} -{\bf q}_{0}) $, where 
$\Omega_{\rm env}$  is the environment volume~\cite{caveat0}. 
The conductance then reads
 \begin{eqnarray}\label{sc-cond}
g &=&  \left(4\pi^2\,\Omega_{\rm env} \right)^{-1} 
 \int_{0}^{\infty}\! \! {\rm d} t  \; {\rm d} t^{\prime} 
 \int_{\rm \Omega_{env}} \!\!{\rm d}  {\bf q} _{0} \; {\rm d} {\bf q}
 \int_{\rm L}\!\!{\rm d} {\bf y}_{0}   \int_{\rm R} \!{\rm d} {\bf y}
  \nonumber   \\&\times&
   \sum_{ \gamma, \Gamma ; \gamma^{\prime}, \Gamma^{\prime}}
    \left( C_{\gamma} \; C_{\Gamma}
   \; C_{\gamma^{\prime}}\;C_{\Gamma^{\prime}}\right)^{\frac{1}{2}} \;\;
   e^{ i (\Phi_{\rm sys} + \Phi_{\rm env} +\Phi_{\cal U})}.\qquad
\end{eqnarray}
This is a quadruple sum over classical paths of the
system ($\gamma$ and $\gamma'$, going from ${\bf y}_0$ to ${\bf y}$)
and the environment ($\Gamma$ and $\Gamma'$, going from ${\bf q}_0$ 
to ${\bf q}$), with action phases
$\Phi_{\rm sys} = 
S_\gamma({\bf y}_0,{\bf y}; t)- S_{\gamma'}({\bf y}_0,{\bf y}; t')$,
$\Phi_{\rm env} = 
S_\Gamma ({\bf q}_0,{\bf q}; t)- S_{\Gamma'}({\bf q}_0,{\bf q}; t')$
and 
$ \Phi_{\cal U} =  
{\cal S}_{\gamma,\Gamma } ({\bf y}_0,{\bf y} ;{\bf q}_0,{\bf q};t)
- {\cal S} _{\gamma',\Gamma'}({\bf y}_0,{\bf y} ;{\bf q}_0,{\bf q};t')$.  
We are interested in the conductance averaged over energy variations, 
and hence
look for contributions  to Eq.~\eqref{sc-cond} with stationary 
$\Phi_{\rm sys},\Phi_{\rm env}$.
The first such contributions are the diagonal ones
with $\gamma=\gamma^{\prime} $ and $\Gamma=\Gamma^{\prime} $, for which 
$\Phi_{\cal U}=0$. They are ${\cal U}$-independent and
give the classical, Drude conductance,
$g^{\rm D} =  N_{\rm L}N_{\rm R}/( N_{\rm L} +N_{\rm R})$. The leading 
order correction to this comes from weak-localization paths $\gamma$ and 
$\gamma'$~\cite{Ale96,wl_semicl,Jac06,Bro06-cbs}
(see Fig.~\ref{fig:weakloc}), with $\Gamma'=\Gamma$ for environment paths
\cite{caveat1}.
In the absence of dephasing these
contributions accumulate a phase difference 
$\delta \Phi_{\rm sys}$. Semiclassically these contributions give 
$g_0^{\rm wl} 
= - \exp[-\tau_{\rm E}^{\rm cl} / \tau_{\rm D}] \;
N_{\rm L} N_{\rm R}/(N_{\rm L} +N_{\rm R})^2$~\cite{wl_semicl,Jac06,Bro06-cbs}.

In the presence of an environment, 
each weak localization pair of 
paths accumulates an additional 
action phase difference $\delta\Phi_{\cal U}$,
which is averaged over. 
Dephasing occurs mostly in the loop,
when the paths are more than 
the correlation length $\xi$ apart (see Fig.~\ref{fig:weakloc}).  
Thus we can  define
$T_{\xi}=\lambda^{-1} \ln [(\xi /\epsilon L)^2]$
as twice the time between the encounter 
and  the start of dephasing.
If $\xi < \epsilon L$, dephasing
starts before the paths reach the encounter, $T_{\xi}<0$~\cite{caveat2}.
Using the central limit theorem and
assuming a fast decaying interaction correlator,
$\left\langle {\cal U}[{\bf y}_{\gamma}(\tau) , {\bf q}_{\Gamma}(\tau) ] {\cal U}[{\bf y}_{\gamma}(\tau^{\prime}) , {\bf q}_{\Gamma}(\tau') ]  \right\rangle_\Gamma$$\,
 \propto\,$$ \tau_{\phi}^{-1}  \exp[-\{  {\bf y}_{\gamma}(\tau)- {\bf y}_{\gamma}(\tau^{\prime})\}/\xi]$, 
the average phase difference due to ${\cal U}$ 
reads
\begin{eqnarray}\label{detphiu}
\big\langle e^{{\it i} \delta \Phi_{\cal U }} \big\rangle =
 e^{-\frac{1}{2} \langle\delta \Phi^2_{\cal U}\rangle}=
 \exp\left[ - (t_2-t_1-T_{\xi})/\tau_{\phi}\right],
\end{eqnarray}
where $t_1$ ($t_2$) gives the start (end) of the loop.
The derivation then proceeds as for ${\cal U}=0$~\cite{Jac06},
except that during the time $(t_2-t_1-T_{\xi})$,   
the dwell time is effectively divided by 
$[1+\tau_{\rm D}/\tau_\phi]$, 
so the $(t_2-t_1)$-integral generates an extra prefactor of
$[1+\tau_{\rm D}/\tau_\phi]^{-1} \exp[-\tau_{\xi}/\tau_\phi]$
where $\tau_{\xi}=\lambda^{-1} \ln [(L/\xi)^2]$. 
Thus 
the weak localization correction is
as in Eq.~(\ref{eq:gwl-bothcases}) with $\tilde \tau$ given by 
$\tau_{\xi}$.
In contrast to Ref.~\cite{Ale96}, the exponent depends 
on $\xi$ not $\lambda_{\rm F}$.

A calculation of coherent-backscattering with dephasing to be
presented elsewhere enables us to show that our approach
is probability- and thus current-conserving. 
We also point out that (for $\tau_\phi \sim \tau_D$) 
one can ignore the modifications of the classical paths due to
the coupling to the environment, 
as long as $\xi \gg   [\lambda_{\rm F}L / \lambda \tau_D]^{1/2}$.
Thus our method is applicable for
$\xi$ smaller (as well as larger) than
the encounter size $[\lambda_{\rm F}L]^{1/2}$, 
but {\it not} for $\xi \sim \lambda_{\rm F}$. 

{\bf Dephasing lead model.}
We next add a third lead to an otherwise closed dot 
(as in Fig.~\ref{fig:two_models}b), 
and tune the potential on this lead such that the net current through 
it is zero.
Thus every electron that leaves through lead 3 
is replaced by one with an unrelated phase, leading to
a loss of phase information without loss of particles.
In this situation the conductance from L to R is given by 
$g = T_{\rm LR} + T_{\rm L3}\,T_{\rm R3}\;
(T_{\rm L3}+T_{\rm R3 })^{-1}$~\cite{But86}, 
where $T_{nm}$ is the conductance from lead 
$m$ to lead $n$ when we do not tune the potential on
lead $3$ to ensure zero current.
We next note that 
$T_{nm}=  T_{nm}^{\rm D}+ \delta T_{nm} 
+ {\cal O}[N^{-1}]$
where the Drude  contribution,  $T_{nm}^{\rm D}$, is ${\cal O}[N]$ 
and the weak localization contribution, 
$\delta T_{nm}$, is ${\cal O}[1]$. 
If we now expand $g$ for large $N$ and collect all  ${\cal O}[1]$-terms
we get
\begin{equation}
\label{3leadmodelWL}
g^{\rm wl}  
= \delta T _{\rm LR} +  
\frac{ 
(T^{\rm D}_{\rm L 3})^2 \; \delta T_{\rm R 3}
+(T^{\rm D}_{\rm R 3})^2 \; \delta T_{\rm L 3} }
{ (T^{\rm D}_{\rm L 3} +T^{\rm D} _{\rm R 3})^2}.    
\end{equation}
For a cavity perfectly connected to all three leads
(with $W_{L,R,3} \gg \lambda_{\rm F}$), the 
Drude and weak localization results for ${\cal U}=0$
(at finite-$\tEc$) ~\cite{Jac06} can be substituted
into Eq.~(\ref{3leadmodelWL}), immediately giving
Eq.~(\ref{eq:gwl-bothcases}) with $\tilde \tau$ given by
$\tEc$.

\begin{figure}
\begin{center}
\includegraphics[width=8.3cm]{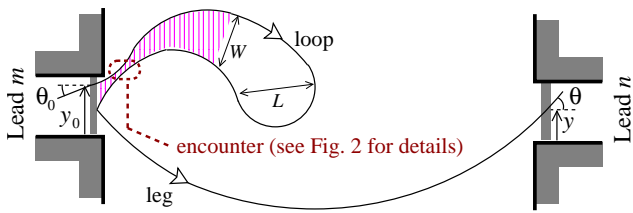} 
\caption{\label{fig:failed-cbs} 
(color online)
A failed coherent-backscattering contribution to
conductance, $\delta T^{\rm cbs}_{nm}$.
It involves paths which return to close but 
anti-parallel to themselves at lead $m$,
but reflects off the tunnel-barrier, remaining in the cavity 
to finally escape via lead $n$. The cross-hatched region is when
the two solid paths are paired (within $W$ of each other).}
\end{center}
\end{figure}

To connect with the numerics of Ref.~\cite{Two04}, we now consider a tunnel-barrier with finite 
transparency $0 \le \rho_3 \le 1$ between the cavity and the dephasing lead.
Introducing tunnel-barriers into the trajectory-based
theory of weak localization is detailed in Ref.~\cite{Whi06}.
It requires three main changes to the theory 
in Ref.~\cite{Jac06}.
(i) The dwell time (single path survival time) becomes
$\tau_{\rm D1}^{-1} = (\tau_0L)^{-1} \sum_m \rho_m W_m$.
(ii) The paired path survival time (for two paths closer than the lead width)
is no longer equal to the dwell time, instead it is
$\tau_{\rm D2}^{-1} = (\tau_0L)^{-1} \sum_m \rho_m (2-\rho_m)W_m$
because survival requires that both paths hitting a tunnel-barrier are
reflected~\cite{Whi06}.
(iii) The coherent-backscattering peak contributes to transmission as 
well as reflection, see Fig.~\ref{fig:failed-cbs}.

For the Drude conductance we need only (i) above,
giving us 
$T^{\rm D}_{nm} = \rho_m \; \rho_n N_mN_n/{\mathcal N}$, where
${\mathcal N}= \sum_k \rho_k N_k$.
For the conventional weak localization correction
we need (i) and (ii).  The contribution's classical path
stays within $W$ of itself for a time $T_W(\epsilon)/2$ on 
either side of the encounter (dashed region in Figs.~\ref{fig:weakloc}
and \ref{fig:failed-cbs}), 
thus we must use the paired-paths survival time,
$\tau_{\rm D2}$, for these parts of the path.  Elsewhere the survival time is
given by $\tau_{\rm D1}$. 
We follow the derivation in Ref.~\cite{Jac06} with these new ingredients,
and the conventional weak localization correction becomes
$\delta T^{(0)}_{nm} 
= -(\rho_m\rho_n N_mN_n/{\mathcal N}^2) (\tau_{\rm D1}/ \tau_{\rm D2})
\exp[-\Theta]$. The exponential with 
$\Theta= \tau_{\rm E}^{\rm op }/\tau_{\rm D_2} + 
(\tau_{\rm E}^{\rm cl } -\tau_{\rm E}^{\rm op } )/\tau_{\rm D_1}$, 
is the probability that the path segments survive
a time $\tEo$ as a pair ($\tEo/2$ either side of the encounter) 
and survive an additional time $(\tEc-\tEo)$ unpaired (to form
a loop of length $\tEc$).
However, 
we must include point (iii) above and consider
the {\it failed coherent-backscattering}, 
shown in Fig.~\ref{fig:failed-cbs}. 
We perform the backscattering calculation following Ref.~\cite{Jac06}
(see also~\cite{Bro06-cbs}) but using $\tau_{\rm D2}$
when the paths are closer than $W$ and $\tau_{\rm D1}$ elsewhere.   
We then multiply the result by the probability that 
the path reflects off lead $m$ and then escapes through lead $n$.
This gives a contribution,
$\delta T^{\rm cbs}_{nm} 
= -(\rho_m(1-\rho_m)\rho_n N_mN_n/{\mathcal N}^2)
\exp[-\Theta]$
assuming $n\neq m$.  There is a second such contribution with 
$m \leftrightarrow n$.
Summing the contributions for $m\neq n$
\begin{eqnarray}
\delta T_{nm} 
&=& (\rho_m\rho_n N_mN_n/{\mathcal N}^2)   
(\rho_m+ \rho_n  - \tilde{\mathcal N }/{\mathcal N})  
{\rm e}^{-\Theta}
\quad \label{gnmwl}
\end{eqnarray}
where $\tilde{\mathcal N }= \sum_k \rho_k^2N_k$.
If only the dephasing lead has a tunnel-barrier,
substituting the Drude and weak localization results 
into Eq.~(\ref{3leadmodelWL}), we obtain
Eq.~(\ref{eq:gwl-bothcases}) with $\tilde \tau=
(1- \rho)\tau_{\rm E}^{\rm op} + \tau_{\rm E}^{\rm cl}$. 
In this case the exponential in Eq.~(\ref{eq:gwl-bothcases}) is
the probability that a path 
does not escape into the dephasing lead in either the paired-region
or the extra time $(\tEc-\tEo)$ unpaired (for the loop to form).

To generalize our results to $j$ dephasing leads,  
we expand the relevant conductance formula in powers of
$N$ and collect the ${\cal O}[1]$-terms. Then 
$g^{\rm wl} = \delta T_{\rm LR} + \sum_{m=1}^j 
(A_m \delta T_{{\rm L}m} 
+ B_m \delta T_{{\rm R}m})$,
where the sum is over all dephasing leads.
The prefactors $A_m,B_m$ are combinations of Drude conductances and thus
independent of $\tEc,\tEo$, we need them to get power-law dephasing.
However we can already see that there must be 
exponential decay with $[(1- \rho)\tEo + \tEc]/\tau_\phi$,
as for $j=1$, 
where now
$\tau_\phi^{-1}= (\tau_0L)^{-1}\sum_m \rho_m W_m$ 
and 
$\rho \tau_\phi^{-1} = (\tau_0L)^{-1}\sum_m \rho_m^2 W_m$.

{\bf Conclusions.} We first observe that the dephasing-lead model
has no independent parameter $\xi$. To our surprise it is the Fermi 
wavelength, not the dephasing-lead's width, which plays a role similar 
to $\xi$. Thus a dephasing-lead model cannot mimic 
a system-environment model with $\xi\sim L$ at finite $\tEc$.
Our second observation is that Eq.~(\ref{eq:gwl-bothcases}) with
$\tilde\tau= \tau_{\xi}$ is for a regime where ${\cal U}$ does not
affect the momentum/energy of classical paths. Therefore, it does not
contradict the result with $\tilde\tau = \tEc$ in Ref.~\cite{Ale96},
valid for $\xi$ so small that dephasing occurs via a
``single inelastic process with large energy transfer''~\cite{caveat}.
Intriguingly their result is similar to
ours for the dephasing-lead model.
Could this be due to the destruction of classical determinism
by the dephasing process in both cases?

We finally note that conductance fluctuations (CFs)
in the dephasing-lead model exhibit an exponential dependence
$\propto \exp[-2 \tau_{\rm E}/\tau_\phi]$ for $\rho_3 \ll 1$~\cite{Two04},
but recover the universal behavior of 
Eq.~(\ref{eq:dephasing-without-Ehrenfest}) for 
$\rho_3=1$~\cite{comment-UCFs}. 
However external noise can lead 
to dephasing of CFs 
with a $\tau_{\rm E}$-independent exponential term \cite{comment-UCFs},   
similar to the one found above for weak-localization 
in the system-environment model.
Thus our conclusion, that dephasing is system-dependent in 
the deep semiclassical limit, also applies to conductance fluctuations.

We thank I. Aleiner, P. Brouwer, M. B\"uttiker and M. Polianski
for useful and stimulating discussions.
The Swiss National Science Foundation supported CP, and funded
the visits to Geneva of PJ and RW.
Part of this work was done at the Aspen Center for Physics.


\begin{thebibliography}{99}
\bibitem{caveat}
Ref.~\cite{Ale96} mentions that they do not expect their result
to hold when $\xi$ is a classical scale.
\end{thebibliography}
\end{document}